\documentclass{article}
\usepackage[english]{babel}
\usepackage[letterpaper,top=2cm,bottom=2cm,left=3cm,right=3cm,marginparwidth=1.75cm]{geometry}

\usepackage{amsmath}
\usepackage{graphicx}
\usepackage{color}
\usepackage[colorlinks=true, allcolors=blue]{hyperref}
\usepackage{authblk}

\usepackage{xspace}

\newcommand{\covid}{COVID-19\xspace}
\def\cpp{{\tt C++}\xspace}

\usepackage[style=numeric-comp,sorting=none]{biblatex}
\begin{document}

\begin{titlepage}

\vspace*{-1.5cm}
\centerline{\large THE HEP SOFTWARE FOUNDATION (HSF)}
\vspace*{1.5cm}
\noindent
\begin{tabular*}{\linewidth}{lc@{\extracolsep{\fill}}r@{\extracolsep{0pt}}}

\\
 & & HSF-TN-2023-01 \\  
 & & October 2023 \\ 
 & & \\
\end{tabular*}

\vspace*{4.0cm}

{\bf\boldmath\huge
\begin{center}
  Training and Onboarding initiatives in High Energy Physics experiments
\end{center}
}

\vspace*{2.0cm}

\begin{center}
S. Hageb\"ock$^1$, A. Reinsvold Hall$^2$, N. Skidmore$^3$, G. A. Stewart$^1$, G. Benelli$^4$, B. Carlson$^{5,6}$, C. David$^7$, J. Davies$^8$, W. Deconinck$^9$, D. DeMuth, Jr.$^{10}$, P. Elmer$^{11}$, R. B. Garg$^{12}$, K. Lieret$^{11}$, V. Lukashenko$^{13,14}$, S. Malik$^{15}$, A. Morris$^{16}$, H. Schellman$^{17}$, J. Veatch$^{18}$, M. Hernandez Villanueva$^{19}$
\bigskip\\
{\it\footnotesize
$ ^1$European Organization for Nuclear Research (CERN), Geneva, Switzerland, 
$^2$United States Naval Academy, Annapolis, MD, USA,
$^3$University of Warwick, Coventry, United Kingdom,
$^4$Brown University, Providence, Rhode Island, USA,
$^5$Westmont College, Santa Barbara, USA,
$^6$University of Pittsburgh, Pennsylvania, USA,
$^7$African Institute for Mathematical Sciences, South Africa,
$^8$University of Manchester, Schuster Building, Manchester, UK,
$^9$University of Manitoba, Winnipeg, Canada,
$^{10}$Valley City State University, North Dakota, USA,
$^{11}$Princeton University, Princeton, New Jersey, USA,
$^{12}$Stanford University, Stanford CA, USA,
$^{13}$Nikhef National Institute for Subatomic Physics, Amsterdam, Netherlands,
$^{14}$Institute for Nuclear Research,  National Academy of Sciences of Ukraine, Kyiv, Ukraine,
$^{15}$University of Puerto Rico, Mayaguez, Puerto Rico,
$^{16}$Aix Marseille Univ, CNRS/IN2P3, CPPM, Marseille, France,
$^{17}$Oregon State University, Corvallis, Oregon, USA,
$^{18}$California State University, Long Beach, California, USA,
$^{19}$Deutsches Elektronen-Synchrotron DESY, Hamburg and Zeuthen, Germany
}
\end{center}

\vspace{\fill}

\begin{abstract}
  \noindent
  In this paper we document the current analysis software training and onboarding activities in several High Energy Physics (HEP) experiments: ATLAS, CMS, LHCb, Belle II and DUNE. Fast and efficient onboarding of new collaboration members is increasingly important for HEP experiments as analyses and the related software become ever more complex with growing datasets. A meeting series was held by the HEP Software Foundation (HSF) in 2022 for experiments to showcase their initiatives. Here we document and analyse these in an attempt to determine a set of key considerations for future experiments.
\end{abstract}

\vspace*{2.0cm}

\vspace{\fill}

{\footnotesize
\centerline{\copyright~Licence \href{http://creativecommons.org/licenses/by/4.0/}{CC-BY-4.0}.}}
\vspace*{2mm}


\end{titlepage}

\clearpage

\section{Introduction}\label{introduction}

Onboarding refers to the process of integrating new members into an organisation and providing them with the knowledge and skills to become effective members of said organisation. The software used in High Energy Physics (HEP) experiments is becoming increasingly complex and challenging for new members to gain proficiency with. Moreover, with collaborations often consisting of more than 1000 members, induction into the collaborations' day-to-day activities and inner structure is an equally important part of the onboarding process for HEP experiments. It is therefore important that experiments have effective and sustainable software training and induction programs. Many details of the analysis tools and software frameworks are experiment-specific, but there is value in comparing training events run by different collaborations and seeing the variety of solutions to common challenges.
Following an initial workshop in 2020~\cite{hsftraining_challenge},
a meeting series~\cite{DAWG_training_indico1, DAWG_training_indico2} was held by the HEP Software Foundation (HSF)~\cite{hsf}  in 2022 for experiments to showcase their initiatives. Preceeding this, DIANA-HEP~\cite{diana-hep}, IRIS-HEP~\cite{iris-hep} and the HSF organized workshops~\cite{wlcg_napoli,how_jlab} bringing together the training experiences and needs of different experiments to produce a common software curriculum. This reflected the growing consensus in the HEP community on the need for training programs to bring researchers up to date with new software technologies, as expressed in the HSF Community White Paper~\cite{cwp_training}. 

In this paper, Section \ref{experiments} describes how training and induction activities are organised in individual experiments. Section \ref{analysis} analyses these initiatives and discusses the general themes, and Section \ref{bestpractices} outlines key considerations for future experiments when creating training and induction events and materials.
 
\section{Experiment approaches}\label{experiments}

The meeting series hosted by the HSF Data Analysis working group brought together several experiments to present and discuss their software training and induction initiatives for new collaboration members. Each experiment's approach is described below.



\subsection{ATLAS}\label{atlas}

A Toroidal LHC ApparatuS, nicknamed the ATLAS experiment~\cite{ATLAS_Collaboration}, is situated at point 1 of CERN's Large Hadron Collider (LHC). Like its sister experiment CMS, ATLAS is a general-purpose detector dedicated to investigating a diverse array of physics topics. This includes precision measurements within the Standard Model (SM) with a prime focus on the properties of the Higgs boson and searches for Beyond-the-Standard-Model (BSM) physics. ATLAS boasts an exceptionally active and devoted physics program dedicated to both proton physics and heavy-ion physics. With over 6000 members and 3000 scientific authors, hailing from 182 institutions across 42 countries~\cite{ATLAS_structure}, ATLAS represents one of the most ambitious and cooperative experiments ever realised in the history of scientific research.  
\subsubsection{Training initiatives}
The ATLAS collaboration has been organizing specialized training for analysis software and workflows since 2004. However, these events became more comprehensive and regularized with the start of the LHC in 2008. The training programs are carefully designed to cater to both new members of the collaboration and experienced individuals seeking deeper insights.
Up until 2020, these training events were conducted in-person at CERN, without any remote connection. However, the outbreak of the \covid pandemic prompted a paradigm shift, leading ATLAS to  transform their software tutorials from in-person to remote sessions, featuring pre-recorded lectures and interactive question-answer sessions.
This approach proved to be highly successful as it enabled participants from all corners of the globe to access the valuable training resources and therefore ATLAS decided to continue the remote sessions. Today, ATLAS employs a variety of different tutorials to effectively educate its members, as detailed below.\\\\
\textbf{ATLAS Induction day \& Software Tutorial:} This comprehensive week-long tutorial comprises two main segments: the first day, known as the induction day, serves to introduce participants to the ATLAS experiment and provides a broader perspective of working in a HEP research environment. A part of the induction day is reserved to guide students through the process of setting up their CERN and ATLAS computing accounts, making this complex process easier for them. 
The subsequent four days are wholly devoted to data analysis software 
training.

In a recent development, the software tutorial underwent a structural transformation aimed at enhancing its pedagogical value and adopting a project-based approach where participants are guided through the process of conducting an end-to-end analysis. The current implementation of this new framework involves the application of the $2^{\textrm{nd}}$ generation LeptoQuark (LQ) analysis on ATLAS Run 2 data~\cite{LQ_analysis}. Throughout the tutorial week, students are expertly guided through each step of the analysis using a combination of presentations and interactive hands-on-tutorials. Wherever possible, code is provided in the form of {\tt Jupyter} notebooks \cite{Kluyver2016jupyter}, streamlining the processes of code editing and execution. Emphasis is placed on team-building and networking, achieved by assembling participants into small groups along with a mentor who all work together throughout the four days.
The first trial of this new format was performed in September 2022 at SLAC~\cite{ATLAS_SLAC_tutorial}, yielding remarkable success and great feedback from the students. Currently, this tutorial is  organized quarterly at CERN, offering back-to-back sessions for in-person and remote participation and annually in the USA for in-person participants.

ATLAS also maintains a well-structured software documentation~\cite{ATLAS_SoftDocs}, which includes user-friendly guides for essential tools like {\tt Git} and {\tt CMake}, along with a comprehensive guide for the ATLAS software {\tt Athena} \cite{atlascollaboration_2019}.
Moreover, to facilitate learning at one's own pace, ATLAS provides a self-guided software tutorial~\cite{ATLAS_selftutorial} that remains readily accessible to students whenever they require it.
\\\\
\textbf{ATLAS Software Development Tutorial:} This tutorial~\cite{ATLAS_SoftDevTutorial} is designed to further advance the skill set of students who have already completed the software induction tutorial and possess a solid foundation in programming languages such as {\tt \cpp}, {\tt Python}, and the ATLAS software {\tt Athena}. The core objective of this tutorial is to enhance participants' proficiency in writing high-quality, error-free, and sustainable code. The curriculum includes essential yet complex subjects like multithreading, databases, {\tt Git} version control, code debugging, and more. Through a blend of presentations and hands-on exercises, participants are equipped with practical skills and theoretical knowledge, fostering a comprehensive learning experience. 
\\\\
\textbf{A year in ATLAS:} Organized by the ATLAS Early Career Scientist board~\cite{ATLAS_ECSB}, this event caters to the students who have already completed their first year in the ATLAS collaboration and who have likely finished their qualification projects. During this half-day event, participants learn about the finer intricacies of the ATLAS organization and its mentoring framework. The event also sheds light on specific processes, such as accumulating Operation Task Points (OTPs) and securing speaking opportunities at national and international conferences, through detailed presentations. A question-answer session is held at the end to provide participants with the chance to seek clarifications, gain deeper insights, and engage directly with the experts.
\\\\
\textbf{Other training events and resources:} ATLAS regularly organizes various workshops targeting specific topics such as detector upgrades, tracking mechanisms, flavor-tagging tools etc. There are also numerous active mailing lists and e-groups available that allow individuals to quickly connect with experts for technical support. 

In addition to this, ATLAS is actively engaged with various social media platforms, utilizing them as avenues to disseminate knowledge. Notably, ATLAS shares educational content through its YouTube channel~\cite{ATLAS_socialMedia} where a wealth of videos provide in-depth insights into the ATLAS detector and its underlying physics. This resource not only proves invaluable for new collaboration members seeking knowledge but also serves as a means for active members to deepen their knowledge.
\subsection{CMS}\label{cms}


The Compact Muon Solenoid (CMS) experiment at the CERN LHC is an all-purpose detector that is used to perform precision measurements of the SM, test properties of the Higgs boson, and look for evidence of BSM physics such as dark matter or supersymmetry. 

The CMS Collaboration has over 6000 members, including engineers, undergraduate students, and approximately 2100 Ph.D. physicists and 1200 doctoral students~\cite{CMS_people}. Over 240 institutes participate in the CMS Collaboration, representing over 50 different countries and regions around the globe. 

\subsubsection{Training initiatives}

CMS offers several different training events to introduce new members to the collaboration and teach essential analysis and software development skills. Many of these events emphasise hands-on practice and networking opportunities. 

\paragraph{CMS Induction Course}

To introduce newcomers to the collaboration, CMS offers an Induction Course once or twice a year. This course focuses on opportunities to learn the collaboration structure and gain an overview of the experiment, rather than teaching hands-on programming skills. The course is split into two days and is currently held in hybrid mode, with both in-person and remote options, and is fully recorded. Lectures generally include talks from the experiment spokesperson and physics coordination; overviews of the different aspects of the experiment including detector subsystems, trigger, offline and computing; and an introduction to groups such as the Diversity Office, Communications Office, and the CMS Secretariat. So far, the course has been offered 13 times since 2014.

\paragraph{Data Analysis School}

To learn analysis software, the most complete training offered is the CMS Data Analysis School (DAS)~\cite{LPC_DAS}. This event was started in 2011 and has been offered 27 times so far, typically 2-3 times per year. DAS takes place over an entire week, usually at either the LHC Physics Center (LPC) at Fermilab or at CERN, although other locations (Pisa, Taipei, DESY, Kolkata, Bari, Daegu, Beijing) have also hosted. The school is geared towards new Ph.D. students, but is also attended by some undergraduate students as well as by postdocs or junior faculty without previous CMS analysis experience. The typical attendance is 50 to 70 students. 

The DAS structure emphasizes hands-on exercises and team work. Over the years the format has evolved slightly, but the basic structure is the following:
\begin{itemize}
\item \textbf{Mandatory pre-exercises} are used to get all students set up with the proper computing accounts and familiar with the basics so they are able to hit the ground running once they arrive. The deadline for completion is before the start of the school. 
\item \textbf{Lectures} on the first two days introduce students to key concepts about analysis as well as the collaboration in general. Topics include: CMS Physics, Introduction to the LHC and the CMS Detector, Software/Analysis Tools, Diversity and Inclusion, and Communications/Outreach. 
\item \textbf{Short exercises} are two-hour sessions that cover the essentials of specific objects and basic analysis ingredients, such as muons, jets, tracking, statistics, etc.
\item \textbf{Writers PUB} is led by senior CMS members, who explain the publication process of a CMS paper.
\item \textbf{Long exercises} are the core of the DAS experience. The students are divided into small analysis teams and, with guidance from facilitators, seek to finish a complete physics analysis by the end of the week. 
\item A \textbf{mini-symposium} on the last day wraps up the school. All teams present their  results from the long exercise, and a panel of judges decides on the best presentation. Members of the winning team typically receive a small prize (like an LPC coffee mug) as well as coveted bragging rights. 
\end{itemize}

Networking is an important component throughout the school. Each participant has a commitment and responsibility to their long exercise analysis team (and not to facilitators or
organizers), which is the key to their individual and team success. While all team members participate in the same long exercise, team members are assigned different short exercises so that the team as a whole can have full coverage of the needed tools for the long exercise analysis. All team members participate in the slides preparation, and all team members are required to present during the final mini-symposium at the end. 

DAS is viewed as an essential data analysis ``boot camp'', but its success requires a major effort from many people. There are typically close to 50 facilitators per DAS, including many postdocs who are responsible for leading the short and long exercises. At the LPC, many of the facilitators are part of the LPC Distinguished Researcher program~\cite{LPC_DR}, which provides funding for postdocs and scientists/faculty to stay at Fermilab and contribute to the LPC community by leading training events, physics analysis, and hardware or software efforts. Many postdocs and senior graduate students once participated in DAS as students and now choose to serve as facilitators. 
Two full-time LPC support staff members, Gabriele Benelli and Marguerite Tonjes, the two LPC co-coordinators (who serve two-year appointments), and the CMS Schools Committee guide the overall organization. 

To adapt to the \covid pandemic, the LPC hosted a virtual DAS in January 2021 and 2022. Instead of one week, the event was spread out over two weeks to mitigate the fatigue from spending too many hours videoconferencing per day. The virtual DAS was still successful in teaching core analysis skills, but it was harder to maintain the crucial networking aspect of the school. Fortunately, in January 2023 the LPC was able to once again host DAS in-person.

\paragraph{Hands-On Tutorial Sessions}
For training on specific topics, the Fermilab LPC offers Hands-On Tutorial Sessions (HATS)~\cite{LPC_HATS} during the spring and summer each year. There are typically about 15 to 20 HATS per year focused on specific physics objects or tools for data analysis such as {\tt Git}, machine learning, the CMS trigger system, jet algorithms, or scientific {\tt Python}. Each tutorial includes both lectures and hands-on exercises and typically takes place over one or two afternoons. Participants are encouraged to attend in-person at Fermilab, but remote participation via Zoom and Mattermost discussions are also possible. The tutorials are all recorded, with the videos and transcripts posted to the CERN Document Server to serve as useful reference materials. Recently, there have been efforts to move the HATS to a software carpentry model \cite{wilson-software-carpentry} to simplify maintenance.

\paragraph{Other training events}
Additional workshops or schools are also offered as necessary to provide training on specific topics. In recent years, these have included CMS Upgrade Schools to train new contributors to the HL-LHC upgrades and dedicated workshops or hackathons led by Physics Object Groups (POGs) or Detector Performance Groups (DPGs) to prepare for data taking or after the release of new tools, etc. A general CMS Physics Object School (CMSPOS) has been offered twice. CMSPOS adopted a similar structure to DAS, but with a focus on learning about physics objects such as electrons or muons in more detail. The goal of CMSPOS is to train people who will become developers for the software of particular POGs or DPGs.

The LPC regularly offers graduate-level advanced courses, which are organized in association with US universities so students have the option to get course credit. These courses include hybrid (in-person or remote) lectures with homework and exams and have covered a wide range of analysis-related topics, including computational physics, statistics, detectors, and machine learning. Weekly Physics Analysis Discussions hosted remotely and in-person by the LPC provide an informal way for analysers to ask for help and hear about the latest updates and recommendations in CMS analyses. 

\subsection{LHCb}\label{lhcb}
The LHCb detector \cite{alves2008lhcb} is a single-arm forward spectrometer located at point 8 of the LHC \cite{2008evans}. 
It has a broad physics program covering $C\!P$-violation, rare decays, electroweak physics, and more, with often world-leading results.
Currently, the collaboration consists of approximately 1600 members from 96 institutes in 21 countries. Since 2020, this represents an increase of approximately 100 new members. In terms of scientists, 28\% of the members are Ph.D. students, 16\% are post-doctoral researchers, and a further 27\% are at a senior level. These ratios are different for technical members.

\subsubsection{Training initiatives}
The approach to training and onboarding at LHCb is typically community-driven by volunteers, typically Ph.D. students, with limited oversight from academics. The volunteers will organize a section of the onboarding process (typically workshops or talks, detailed below), before finding the next set of volunteers at the end of their tenure.

This highly collaborative peer-to-peer system comes with several marked advantages. Firstly, fresh eyes applied to the training materials each year helps in keeping them up-to-date and relevant. Secondly, the needs of newcomers are well-known to Ph.D. students, who are in a good position to judge the appropriate level of prior knowledge that may be assumed of newcomers. Thirdly, due to these events being held in person wherever possible, they also serve a social function, allowing students to meet and network with other LHCb members, improving the cohesion of early-career scientists within the collaboration. Finally, the prospect of asking questions is far less daunting when posed to more senior peers than to well-established figures within the collaboration.

\paragraph{Starterkit}
Starterkit is an annual introductory hands-on workshop for LHCb newcomers, \cite{Starterkit_website}, \cite{Puig_2017}.
The workshop was established in 2015 by Tim Head, Kevin Dungs, and Albert Puig Navarro with the help of Igor Babuschkin. It was held exclusively in person until 2020, online in 2020-2021, and in hybrid mode in 2022.
Approximately 40 students participate in the in-person training at CERN each year, and about 100 participate online.
The Starterkit is usually organized and taught by Ph.D. students and postdocs.
Before the workshop, teaching tips and tricks are shared amongst volunteers based on experience from the previous year.
Lectures are maintained by the LHCb community, and teachers are required to review and update lecture material prior to their lessons. 
The Starterkit is often the first collaboration event for participants; therefore, the spokesperson, physics coordinator, and early career, gender, and diversity officers give a short introduction to LHCb.

The week-long Starterkit program consists of two parts and assumes no prior knowledge.
The first part follows HSF analysis essentials on basic programming skills \cite{analysis_essentials}, and the second part covers the``First Analysis Steps", which introduces LHCb-specific software \cite{Starterkit_lectures}.
HSF analysis essentials includes hands-on lessons on {\tt Bash}, {\tt Python} (including selected {\tt PyHEP} packages \cite{pyHEP}), {\tt GitLab} \cite{gitLab}, and {\tt Snakemake} \cite{snakemake}.
The``First Analysis Steps" section describes the LHCb dataflow, distributed analysis software, and LHCb-specific reconstruction software, as well as how to best find help at LHCb in a logical order.
The last day is dedicated to a practice session, where participants use their newly acquired skills to process a subset of the LHCb dataset, just as would be done in future analysis work.
In 2017-2019, the first part of Starterkit was shared with the ALICE and SHiP collaborations \cite{refId0}.

Materials from the Starterkit lectures are available all year long for self-study on a public website and serve as a useful starting point for those looking to implement unfamiliar techniques.
This is also handy for analysts interested in open data, as the Starterkit lessons provide easy-to-follow data processing instructions.
Lectures are also recorded and uploaded to the CERN video service \cite{cds}.
To adjust the material each year, prior to the workshop, participants are asked to fill in a quiz, which helps teachers to understand their level.
After the workshop, participants provide feedback on the teaching and organization, which is used to revise the workshop further. The power of Starterkit is its collaborative spirit, peer-to-peer model, clear explanations of collaboration jargon, and annual reviews.
Started by a group of enthusiasts, Starterkit is recognized by many as compulsory Ph.D. training.

\paragraph{Impactkit}
In spring 2016, a follow-on event from the Starterkit, called the Impactkit, was organized by Albert Puig Navarro and Lennaert Bel with the aim of building upon the skills and techniques learned in the Starterkit and introducing more specialized topics. Subsequent Impactkits have been held in one form or another every year since and cover topics such as simulation with {\tt Gauss} \cite{Gauss}, the software trigger with {\tt Moore} \cite{MooreDocs}, particle combining and filtering, and more sophisticated techniques for distributed analysis. In a similar manner to the Starterkit, the last day is typically reserved for a hackathon-style practical session, where participants are able to consolidate what they've learned by working on more open-ended problems. Slightly more compact than the Starterkit, the Impactkit typically takes place over three days, and attending both events enables participants to know all the software that is regularly used by a significant number of LHCb analysts. The documentation can be found in the same location as the Starterkit materials, under the heading of ``Second Analysis Steps'' \cite{Impactkit_lectures}.

\paragraph{StarterTalk and Theory talks}

StarterTalk is an hour-long lecture on LHCb topics organized by postdocs and held during LHCb collaboration weeks.
Established in 2017 by Giulio Dujany and Vitalii Lisovskyi, StarterTalk invites LHCb experts on LHCb detector, reconstruction, and physics topics to share their expertise with their colleagues in a pedagogical way. This helps to improve members' general understanding of the wide range of activities that the collaboration performs. 
Around 20 lectures had taken place by December 2022, with an average in-person attendance of 50 people. 
All lectures are recorded and uploaded to the CERN video service.

As an extension to the StarterTalks, in 2022, the first series of theory talks were organized ahead of the annual Implications Workshop \cite{Implications} that brings together theorists and LHCb members, helping each side to understand the other's needs better. These talks aimed to improve understanding within the collaboration of the theory that underpins its experimental work.


\paragraph{LHCb-UK student talks}
The LHCb-UK student talks were originally a set of talks organized in 2008 by Jonas Rademacker with the intention of fostering collaboration between LHCb groups in the UK. Since 2009, however, the organization of these talks has been completed by student volunteers from UK institutes, with organizers changing every six months. 
The goal of more recent LHCb-UK student talks has been to explain a single aspect of the LHCb machine or method used in LHCb analyses, which have been found to be more useful for students compared to \textit{e.g.\xspace} presentations of entire analyses.

Contrary to what its name may imply, the LHCb-UK student talks are publicly available to anyone, and all previous talks are archived \cite{LHCbUKstudenttalks}. Furthermore, since the global outbreak of \covid, there has been a renewed effort to record these talks, improving accessibility.

\paragraph{Glossary}
Another very useful resource for new members of the collaboration is the LHCb Glossary \cite{LHCbGlossary}. As in any large organization, there is a great deal of jargon and acronyms that can often be very confusing when someone first joins discussions within the collaboration. This document was started in November 2018 by Eduardo Rodrigues, who focused on content, and Henry Schreiner, who helped to set up the {\tt GitBook}. The aim was that this would be a community effort, and anybody could add any term that was not already listed, and many experts from the wide range of activities that LHCb performs contributed in the early days. In recent times, the rate of new additions has slowed somewhat, but with the advent of Run 3 and the associated new jargon that comes with it, the document will likely see a resurgence in its utility.

\paragraph{Improved documentation}
Historically the documentation of software at LHCb has had room for improvement. To this end, there has been a push recently for better documentation of software, particularly that used in the upgraded LHCb trigger \cite{LHCb-TDR-016}. This is the case with The {\tt Allen} \cite{Aaij:2019zbu,LHCb-TDR-021} and {\tt Moore} software packages used by the upgraded trigger. Similarly to the Starterkit and Impactkit, the documentation is presented as a {\tt Sphinx} {\tt GitBook} \cite{AllenDocs,MooreDocs,Sphinx}.

\subsubsection{Migration to Run 3} 

Starting in 2021, training in LHCb started to slowly migrate towards Run 3 software.
As of December 2022, the annual Starterkit covers topics from both Run 2 and Run 3 software. 
In March 2022, a special online Run 3 Starterkit was organized for everyone in LHCb, with around 200 people in attendance.
Each of the topics was introduced by experts, maintaining its hands-on spirit.
This event formed the basis of the Run 3-related materials to be used in the Starterkit going forward.
When the transition period between Run 2 and Run 3 is over, likely, Starterkit will only cover Run 3 topics. Following this, the Run 1 and Run 2 materials will be retired from the workshop, but the materials and lecture recordings will be kept online for self-guidance.




\subsection{Belle II}\label{belleII}

The Belle II experiment~\cite{Belle-II:2010dht} is a HEP experiment located at the SuperKEKB electron-positron collider in Tsukuba, Japan. It is designed for precision measurements of heavy quark and lepton physics involving $B$ mesons and searches for new physics beyond the Standard Model. 
The Belle II collaboration currently consists of more than 1100 members from more than 300 institutions in 27 nations~\cite{belle2-collab}. A detailed summary of the current training model at Belle II can be found in~\cite{Belle_Lieret_2023}.
\subsubsection{A self-study friendly training paradigm}
Taking the disruption of in-person activities caused by the \covid pandemic as an opportunity to rethink the previous training approaches, the Belle II software training material was remodeled to focus on self-study as the primary training mode.
This reflects the observation that the efficiency and scope of synchronous training events are often not optimal, as the training needs, previous experiences, and schedule preferences diverge among the various new members. In addition, the reliance on slides and other hard-to-maintain material used in many in-person events results in high demand for personpower, duplicated efforts, and frequently outdated resources.

The framework for the new training material was therefore designed to specifically tackle the following challenges:

\begin{itemize}
    \item A high degree of \textbf{maintainability and sustainability}. The most important aspect of this is \textbf{testability}: As far as possible, all examples should be tested against the current version of the Belle II software. This is important because the analyst-facing interface of the main software is still evolving. This also means that the training material should be versioned along with the main software.
    \item \textbf{Connecting resources}: In order to keep the material lean and avoid the duplication of content, the training material should be linked to relevant parts of the technical documentation (API documentation). Teaching newcomers to work with these resources should be an objective in itself.
    \item \textbf{Interactivity}: All lessons should be complemented by exercises with complete solutions and optional hints to adapt to the level of the learners.
\end{itemize}

\subsubsection{An all-in-one solution for documentation and training material}


To tackle these challenges, Belle II chose to directly host the documentation in the repository of the Belle II software framework~\cite{Belle_Moll_2011,Belle_github} and render it via {\tt Sphinx} together with the technical documentation.
Every training module corresponds to one or more restructured text files.
Importantly, {\tt Sphinx} allows the inclusion of code from external source files. It also supports sophisticated ways to include \emph{parts} of source files, which is important when building up larger examples step by step while avoiding duplicated code. Importing source code from external files allows running static code checkers and formatters on the code snippets and makes it easy to include them as unit tests.
Unit testing ensures that no pull request that breaks backward compatibility can be merged without updating the relevant training material.
To spot other issues even outside of events, anonymous feedback is solicited by integrating a small submission form at the bottom of every lesson. The Belle II documentation, including the training material, is publicly available~\cite{Belle_github}.

\subsubsection{Experiences}

The integration of the training material with the Belle II software framework makes the setup more complex than that of a wiki system or a stand-alone documentation system. 
This makes it more challenging to incentivize contributions from members less involved with software development.
However, it offers advantages that might justify the increased technical complexity. As discussed earlier, the setup ensures that all training examples remain functional through coupling with the software and performing unit tests. Furthermore, it allows for seamless linking between technical documentation and tutorials, pointing intermediate learners to in-depth resources without sacrificing clarity.

The setup provides a very comprehensive onboarding experience, covering everything from basic physics knowledge and collaborative tools to software prerequisites and grid job submissions. Its effectiveness has been confirmed by several surveys of the participants of workshops shortly after the restructuring of the material. While some participants wished for additional interactive sessions and lectures, the material was very well received.
In the words of one participant: \textit{``Very solid work regarding the textbook! Congrats! Everything was very clear which significantly minimized the need for guidance (...) Software Prerequisites was the highlight of the workshop as it summarized all the necessary tools that no one really spends time on explaining thoroughly to newcomers.''}
\subsection{DUNE}\label{dune}

The Deep Underground Neutrino Experiment (DUNE) is an international collaboration working to measure $C\!P$ violation in neutrino oscillations and simultaneously observe Supernova burst neutrinos. Additionally, the DUNE detectors will be sensitive to MeV scale solar neutrinos and will be used for rare and exotic BSM searches. DUNE consists of 400 collaborators from over 200 institutions in over 30 countries. Descriptions of the science of DUNE are available for the public on the DUNE Science page~\cite{dunescience}.

\subsubsection{Training Initiatives}

\paragraph{Documentation:} Newly onboarded and longstanding researchers can access DUNE computing resources which are inventoried as a one-stop MediaWorks wiki \cite{dunedoc}, much of which require protected access all DUNE collaborators enjoy. The following main blocks form the top level (entry) of the DUNE Computing wiki: Organization \& Partners, Computing Toolbox, Operations \& Monitoring, Working Groups, Resources, and Getting Started. 
Self-study of documentation is an important initial step to training. As unrecognized topics or words are studied, 
a DUNE-specific glossary~\cite{duneabc} 
was developed as a web-based interface to provide perspective.

\paragraph{Basics Computing Tutorials:} 
While fundamentals such as using the {\tt Unix} Shell
, {\tt Git} and {\tt GitHub}
, and {\tt Python} 
are well covered by HSF 
and IRIS-HEP tutorials 
using the Software Carpentry (SC) lesson templates
, DUNE-specific computing training has been provided as multi-day workshops as a part of each collaboration meeting.  Basics addressed are data storage and management, event reconstruction and simulation using the {\tt art} framework~\cite{Green:2012gv} and {\tt LArSoft} tools~\cite{Snider:2017wjd,larsoft.org} and job submission and monitoring as a way to jump-start event simulation and reconstruction work. Several dozen participants are typical for collaboration tutorial events.

A small cadre of dedicated experts provide instruction utilizing SC lesson templates which are coded in {\tt Markdown}, and who practice active learning techniques such as live-coding and quizzes. 
As {\tt GitHub} hosted materials, {\tt GitHub}-pages are used to render the materials as an elegant and interactive resource~\cite{dunetrain}.
A crucial part of the training, to ensure the ability of participants to actively follow the live exercises, is the pre-training setup \cite{dunesetup} that takes new users through the steps needed to ensure that they have valid computer accounts and can already access DUNE interactive computing resources at Fermilab or CERN. 
New users are invited to join a dedicated DUNE Slack channel to get assistance with setup and the tutorial itself. 

This framework also operates well for asynchronous study with the captured Zoom videos of each session embedded. For those revisiting the lessons, and others who learn autonomously, asynchronous access to the lessons is encouraged.

\paragraph{Support} Ensuring progress on the physics goals of DUNE requires user support in addition to documentation and training.   Slack channels play an important role, particularly for asynchronous just-in-time support. During training events, Google documents are used as ``livedocs'' for real-time responses to questions. Afterwards, livedocs become an archive resource that can be studied. DUNE Computing also uses {\tt GitHub} issues\cite{dunefaq} as a mechanism for frequently asked questions. 

Members of the DUNE Computing Consortium have taken on the responsibility of providing documentation, basics computing tutorials, and user support collaboration-wide. Members are committed to leveraging existing training materials provided by the open communities, and to contributing to the development of new learning modules with broadened appeal.

\section{Common themes and challenges}\label{analysis}

The training and induction initiatives described in this paper all have the same goal: to help new members of each experiment quickly learn the necessary skills to contribute to physics analysis and be efficient, well-integrated members of the collaboration. Some of the common themes and challenges when considering training and induction initiatives are discussed in this section. In Section~\ref{bestpractices}, we summarize this discussion into a list of considerations for future initiatives.

\subsection{Designing training events}




With the advent of the \covid pandemic, many training events shifted from in-person to remote, and currently many are offered in a hybrid format. Remote or hybrid events \textbf{remove barriers} to participation that travel can cause, and therefore allow more members of the collaboration to be trained. Similarly, the Belle II approach of self-study training materials ensures that all collaboration members can utilize the resources whenever they need to. 
One compromise that several experiments have adopted is to \textbf{record training events} so that lectures can be watched by participants who were not able to attend the live events. 

The primary benefit of in-person events is an improved ability to \textbf{network} and connect with other people from the same experiment. 
Often, data analysis schools or training events are the first opportunities that new members have to meet people outside of their own institution and to \textbf{hear directly from experiment leadership}. Knowing the internal structure of the collaboration is an important part of the induction process; LHCb and CMS have dedicated talks from the Spokesperson, Physics Coordinator, diversity officers and secretariats. When planning onboarding events, sufficient opportunities for networking with peers should be included, especially during hybrid or remote events. Grouping participants into teams to work on an example analysis (like what is done during the ATLAS and CMS tutorials) can also help facilitate meaningful networking as well as develop soft skills such as presentation and teamwork skills. If possible, such in-person training and induction events should be scheduled directly before, after or even during collaboration weeks as practiced by DUNE. This can permit more early-career members to attend in person, since travel costs have to be paid only once.

Before live events, it is important to make sure that all participants have the \textbf{necessary prerequisites to participate} in the exercises.
These prerequisites include computing accounts and certificates to use computing infrastructure, but also basic knowledge such as how to use {\tt Git} or the {\tt Unix} shell if these are not covered in the event itself. 
Pre-workshop checklists and dedicated support to fulfill said checklist should be set up to help participants prepare for synchronous events. Pre-workshop quizzes can also be used to assess the level of the incoming group and tailor content and timetabling.
After the event, asking students for feedback can help improve future versions of the event. Having a clear way for participants to provide feedback for self-study training initiatives is also valuable as implemented at Belle II with anonymous feedback submission forms.

\subsection{Learning techniques}

To increase retention of the material and student engagement, training initiatives should be interactive and \textbf{incorporate active learning techniques}. For example, in asynchronous training, this can be done with quizzes and exercises with solutions and hints. For synchronous events, live, hands-on exercises with direct support can complement traditional lectures.  It is important to consider that there will be a range of abilities in any group and the rate at which people grasp new concepts will vary. Therefore extra help should be available where required and, whether synchronous or asynchronous, there should be a sufficient pool of instructors such that students can receive timely help if they are stuck on an exercise.

\subsection{Training materials}



Training materials, even those designed for synchronous events, are often used as  \textbf{important reference materials year-round}. Keeping these materials up-to-date is therefore essential to maximizing their impact. It should be expected that when one modifies experiment code, one also has to update the relevant documentation and tutorials. If the training materials are integrated into the experiment's software framework, like what is done in Belle II, then upkeep can be handled via continuous integration and unit tests.  If there are regular synchronous events, then those are also natural opportunities to check all of the related documentation. The training materials should include links to other useful sources of (perhaps more technical) documentation within the experiment for further self-study. A \textbf{glossary} is a popular self-study resource designed to combat the wealth of jargon used within experiments that the whole collaboration should feel responsible for maintaining. 

LHC experiments have also committed to the CERN Open Data policy \cite{CERN-OPEN-2020-013} of making their data open for general consumption after a fixed period of time. The expected use cases of this Open Data are reinterpretation/reanalysis of physics results, education and outreach, technical and algorithmic development and physics research.  Without good documentation on how to use the necessary (open source) experiment software, the CERN Open Data initiative will not fulfil its purpose. These onboarding software training materials should also then be recognised and used as \textbf{resources towards the successful fulfilment of the CERN Open Data policy} by the LHC experiments. 

\subsection{Sustainability of training initiatives}




\textbf{Keeping training materials up-to-date} and running training events takes a significant amount of personpower, and it is often a challenge to find enough people who have the time, ability, and motivation to make training initiatives successful. There are two complementary solutions to this problem: firstly, attracting new facilitators by emphasizing the benefits and rewards of assisting with training; and secondly, minimizing the required personpower by simplifying events or reducing the duplication of effort.

The first solution is to identify and motivate people to assist with training. 
An important part of this is making sure people's \textbf{contributions to training are publicly acknowledged}. 
The HSF Training Working Group~\cite{hsftraining}, for example, maintains a central community website~\cite{hsftraining_community} listing everyone who has helped with one of their events. 
In CMS, many of the facilitators for the Data Analysis Schools~\cite{LPC_DAS} or Hands-on Tutorial Programs~\cite{LPC_HATS} come from the LPC Distinguished Researchers (DR) program~\cite{LPC_DR}. People selected as DRs gain visibility for their leadership roles within CMS, and there is an explicit expectation that DRs will assist with events at the LPC, including training events. Many experiments also follow the LHCb model, which has established a collaboration culture where participants are encouraged to become facilitators in a future round of the StarterKit. Belle II requires all software developers to contribute to training materials, since updating reference materials is an unavoidable step of the standard workflow. Experiments which have a `credit' system for service work could consider granting official credit for leading software training.

Researchers should also see the material as not just for new-starters but as an up-to-date reference guide for use throughout their involvement in the experiment. Many of the more mature collaboration members use the training material on a regular basis and should also be encouraged to contribute to its development and maintenance. This is particularly true for LHC experiments as we move into the HL-LHC era, when all collaboration members will have to become familiar with new software and tools. Finally, establishing the role of software engineer as a viable career path in HEP can help increase the pool of facilitators, especially if assisting with training is explicitly included in the job description. 

The second solution is to \textbf{run training initiatives as efficiently as possible}, to reduce how many facilitators are needed. In part, this means avoiding duplication of effort and using existing resources as much as possible. 
Common skills such as proficiency with {\tt Python} and {\tt \cpp} should be considered an asset across experiments, and tutorials on these topics should be run as such; LHCb has joined with ALICE and SHiP to deliver such lessons in the past. There are also regular events such as the {\tt PyHEP} workshops and the HEP {\tt \cpp} Course~\cite{HEPCPlusPlus} with hands-on tutorials that, due to their wider nature and larger resource pool, can be more effective than events at collaboration level. 
The same applies to training materials; the HSF has developed the HSF Software training centre with modules on cross-experiment topics which can be used to teach fundamental software and analysis skills \cite{hsftraining}. These are used effectively by some experiments and include, but are not limited to, Machine Learning, {\tt Git}, {\tt Singularity/Docker}, {\tt ROOT} \cite{BRUN199781} and other programming languages. 

The available personpower is also a key consideration when deciding on synchronous or asynchronous training. Asynchronous, self-study training materials such as those developed by Belle II, for example, require much less time than organizing extended live events, although in-person events have their own benefits as previously discussed.    
Another way to reduce development overheads is to make training material \textbf{modular} wherever possible. This helps people quickly access the particular task or ``lesson'' with which they need guidance and also aids in the maintenance of the material as it can be factorised out to different experts. Whole analysis examples (such as those used in the CMS and ATLAS tutorials) are valuable but also require significantly more time to develop and maintain. 

\subsection{Long-lasting support}

Throughout their career, when seeking support, collaboration members must ask well-formed questions in the appropriate communications channel. Asking good questions is a skill; the LHCb Starterkit has a dedicated lesson ``Asking good questions'' to train this. Any training initiative should then also teach people (even senior members of the collaboration) about the support channels available and their remit as well as introduce participants to relevant experts or peers on particular topics.

By providing these skills participants should feel \textbf{empowered to ask questions and seek help} when they run into issues applying the training to real-world analysis tasks. There is a wide variety of methods for providing real-world support, including {\tt Git} issues, Slack or Mattermost channels. It is observed that for more informal channels like Slack and Mattermost there is a lower perceived barrier for newcomers to ask questions.

\subsection{Summary and considerations for Future Experiments}\label{bestpractices}

To summarize this Section we propose a list of considerations for training and induction initiatives:

\begin{itemize}
    \item To improve accessibility, have the option for remote participation or guided self-study for those who are not able to attend in-person events
    \item Provide networking opportunities for new members of the collaboration
    \item Where possible include an introduction to collaboration structure and relevant offices
    \item Plan pre-workshop checklists (eg.\ computing accounts, certificates) with dedicated hands-on support sessions to help students complete the checklist before the event
    \item Consider post-event feedback surveys to develop and adapt events and materials
    \item Follow proven pedagogical practices such as active learning techniques to engage participants and increase retention of the material
    \item Ensure hands-on sessions have a good ratio of instructors to students
    \item Keep training materials up-to-date all year round so they can serve as a useful reference for self-study
    \item Reward and motivate the hard work done by facilitators so that training is viewed as a rewarding and vital task within the collaboration
    \item Encourage senior members of the collaboration to contribute to the upkeep of software training material, since they can also benefit from it as a valuable reference
    \item Take advantage of existing training resources such as the many experiment-agnostic training modules provided by the HSF 
    \item Keep material modular where possible. Whilst whole analysis examples are valuable, modular training material aids maintenance and findability
    \item Introduce participants to channels for future support and provide advice on how to ask good questions
\end{itemize}

Finally, leaders of experiment training events should continue to communicate with their colleagues in other experiments so common lessons, challenges, and solutions can be shared. Improved training initiatives across HEP contributes positively to the physics goals of each experiment, which is beneficial for our entire field.

\section{Conclusions}

This paper documents and reviews the approaches taken to onboarding new members of the ATLAS, CMS, LHCb, Belle II and DUNE collaborations. The HSF, by sharing the initiatives and experiences from a range of HEP experiments, have summarised a set of considerations for onboarding new members into future scientific collaborations. Most importantly, scientific collaborations should incentivize the enormous effort, and recognise the importance of, delivering and maintaining training events/materials. To minimise duplicated efforts, where possible, resources should be shared within the community, and materials provided by groups such as the HSF should be exploited.









\section*{Acknowledgements}
AM acknowledges support of the European Research Council Starting grant ALPACA 101040710. JD acknowledges the support of European Research Council Starting grant  Beauty2Charm 852642. BC acknowledges the support of the National Science Foundation grant 2209370. ATLAS tutorials in the US were supported by the US Department of Energy and the National Science Foundation through the US-ATLAS Operations Program. CMS training activities in the US are supported by the US Department of Energy and the National Science Foundation (NSF PHY-2121686) through the US-CMS Operations Program. HSF/IRIS-HEP training programs are supported by NSF grants OAC-1836650, OAC-1829707 and OAC-1829729.

\sloppy
\raggedright
\clearpage
\printbibliography[title={References},heading=bibintoc]

\clearpage
\end{document}